\documentclass[sort&compress,review]{elsarticle}

\usepackage{mathrsfs}
\usepackage{amsfonts}
\usepackage[colorlinks]{hyperref}
\usepackage{color}

\newtheorem{thm}{Theorem}
\newtheorem{lem}[thm]{Lemma}
\newtheorem{cor}[thm]{Corollary}
\newdefinition{rmk}{Remark}
\newdefinition{df}{Definition}
\newproof{pf}{{\it\noindent{\bf Proof}}}
\def\Q.E.D{\hfill$\blacksquare$}
\newcommand{\dahao}{\fontsize{11pt}{11pt}\selectfont}
\usepackage{amssymb}
\journal{Journal of Computer and System Sciences}

\begin{document}

\begin{frontmatter}



\title{Another approach to the equivalence of measure-many one-way quantum finite automata and its application}


\author[a]{Tianrong Lin}
\ead{tianrong671@yahoo.com.cn}

\address[a]{Dept.~of Comput.~Sci., Fukien University of Technology GM Information College, China\\$\,$\\{\dahao In memory of my grandparents}}

\begin{abstract}
In this paper, we present a much simpler, more direct, and more elegant approach to the equivalence problem for {\it measure-many one-way quantum finite automata} (MM-1QFAs). The approach is essentially a generalization of the work of Carlyle [J.~Math.~Anal.~Appl.~7 (1963) 167--175]. Specifically, we reduce the equivalence problem for MM-1QFAs to the equivalence of two (initial) vectors.

\par
As an application of this approach, we use it to solve the equivalence problem for {\it enhanced one-way quantum finite automata} (E-1QFAs) introduced by Nayak [Proceedings of the 40th Annual IEEE Symposium on Foundations of Computer Science, 1999, pp.~369--376]. We prove that two E-1QFAs $\mathcal{A}_1$ and $\mathcal{A}_2$ over $\Sigma$ are equivalent if and only if they are $(n_1^2 + n_2^2 - 1)$-equivalent, where $n_1$ and $n_2$ are the numbers of states in $\mathcal{A}_1$ and $\mathcal{A}_2$, respectively.\\

\end{abstract}


\end{frontmatter}

% \linenumbers

\section{Introduction}
\label{sec1}

The theory of quantum computing is unquestionably one of the most active and prominent research fields in the theory of computing \cite{1,2,3}. Several models of quantum computation have been developed, such as {\it quantum Turing machines} \cite{5,6}, {\it quantum circuits} \cite{7,8}, and the quantum generalizations of {\it finite automata}, i.e., {\it quantum finite automata} (QFAs) \cite{9,10,11,12,13,14,15,16,22}. In particular, the study of QFAs provides valuable insight into the nature of quantum computation, since QFAs can be viewed as the simplest theoretical model based on quantum mechanics.

The so-called {\it measure-many one-way quantum finite automata} (MM-1QFAs), introduced in \cite{10}, is a type of QFA model whose tape head moves one cell to the right at each computation step, with a measurement performed after every step. Several works have studied the language recognition power of MM-1QFAs \cite{10,11,14,17,18,19,20,21}. Incidentally, the so-called {\it enhanced one-way quantum finite automata} (E-1QFAs) introduced by Nayak \cite{22} can be viewed as a generalization of MM-1QFAs.

Just as with the equivalence problem for classical finite automata \cite{23,24,25,34,35}, the notion of equivalence provides a way to classify MM-1QFAs over the same alphabet. Regarding the equivalence problem for MM-1QFAs, Li and Qiu \cite{26} proved---using the so-called {\it 1QFA with control language} \cite{11}---that two MM-1QFAs $\mathcal{A}_1$ and $\mathcal{A}_2$ over the same alphabet are equivalent if and only if they are $(3n_1^2 + 3n_2^2 - 1)$-equivalent, where $n_1$ and $n_2$ are the numbers of states in $\mathcal{A}_1$ and $\mathcal{A}_2$, respectively, and the factor $3$ comes from the number of states in the minimal DFA \cite{23,24,25} recognizing the regular language $g^*a\{a,g,r\}^*$. Some works have also addressed the equivalence problem for other variants of quantum finite automata \cite{27,28,29,30}. However, the equivalence problem for E-1QFAs remains open. A more comprehensive survey on this topic can be found in \cite{31} by Gruska.

We note that the method used by Li and Qiu \cite{26} for the equivalence problem of MM-1QFAs is somewhat roundabout and complicated. Therefore, the first aim of this paper is to present a much simpler, more direct, and more elegant approach to the equivalence problem for MM-1QFAs. Our motivations are as follows: (1) Different mathematical methods and concepts are often used to investigate the same problem in order to gain a deeper understanding; (2) it is interesting in its own right to find a more general method for the equivalence problem of MM-1QFAs; (3) we wish to determine whether the previous upper bound $3n_1^2 + 3n_2^2 - 1$ can be improved. These considerations lead us to transform the word function of MM-1QFAs (defined in a ``cumulative manner," as described later) into a ``non-cumulative version." We then improve the previous upper bound to $n_1^2 + n_2^2 - 1$ by proving the following theorem.
 
 \begin{thm}\label{thm1}
  Let $\mathcal{A}_i=(Q_i,\{U_i(\sigma)\}_{\sigma\in\Sigma\cup\{\$\}},|\pi_i\rangle,\mathcal{O}_i)$, $i=1,2$, be two MM-1QFAs over $\Sigma$. Then $\mathcal{A}_1$ and $\mathcal{A}_2$ are equivalent if and only if they are $(n_1^2+n_2^2-1)$-equivalent, where $n_1$
 and $n_2$ are the numbers of states in $\mathcal{A}_1$ and $\mathcal{A}_2$, respectively.
\end{thm}

As mentioned earlier, the E-1QFA model \cite{22} can be seen as a finite-memory version of the mixed-state MM-1QFA. Thus, the approach developed for MM-1QFAs can also be applied to E-1QFAs. Accordingly, as our second aim, we apply the above approach to solve the equivalence problem for E-1QFAs (which has remained open) by proving the following theorem.

\begin{thm}\label{thm2}
 Let $\mathcal{A}_i=(Q_i,Q_{acc,i},Q_{rej,i},\{\mathcal{U}^{(i)}_{\sigma})\}_{\sigma\in \Sigma\cup\{\#,\$\}},\rho_i,\mathcal{O}_i)$, $i=1,2$, be two E-1QFAs over $\Sigma$. Then $\mathcal{A}_1$ and $\mathcal{A}_2$ are equivalent if and only if they are $(n_1^2+n_2^2-1)$-equivalent, where $n_1$ and $n_2$ are the numbers of states in $\mathcal{A}_1$
 and $\mathcal{A}_2$, respectively.
\end{thm}

The remainder of the paper is organized as follows. Section \ref{sec:preliminaries} reviews basic concepts and notations. Sections \ref{sec:proof_of_theorem1} and \ref{sec:proof_of_theorem2} are devoted to the proofs of Theorems \ref{thm1} and \ref{thm2}, respectively. Section \ref{sec:conclusions} contains concluding remarks.

\section{Preliminaries}
\label{sec:preliminaries}

For convenience, we briefly review some basic notions needed in the sequel. For a more exhaustive treatment of linear algebra, we refer the reader to \cite{32}. We also refer the reader to \cite{1,2,3} for a thorough treatment of quantum theory.

\subsection{Some notation on Linear algebra}

Let $\mathbb{C}$ denote the field of complex numbers. {\color{blue}Let $M$ be a complex matrix, i.e.,
 $\left(
         \begin{array}{ccc}
           a_{11} & \cdots & a_{1n} \\
           \cdots & \cdots & \cdots \\
           a_{m1} & \cdots & a_{mn} \\
         \end{array}
       \right)$
with $a_{ij}\in \mathbb{C}$ for all $1\leq i\leq m$ and $1\leq j\leq n$.} Sometimes, we use $(a_{ij})_{m\times n}$ to denote the matrix $M$. In particular, $1\times n$ (resp.~$n\times 1$) complex matrices are called $n$-dimensional row vectors (resp.~column vectors). If $m=n$, then $M$ is called a {\it complex square matrix} of order $n$ (or $m$), and is sometimes called an $n$-order (or $m$-order) complex matrix. Let $M=(a_{ij})_{m\times n}$ be an $m\times n$ complex matrix. The transpose of $M$ is denoted by $M'$, i.e., $M'=(a_{ji})_{n\times m}$, and the conjugate transpose of $M$ is denoted by $M^{\dagger}$. In this paper, the set of all $n$-order complex matrices is denoted by $\mathbb{M}_n(\mathbb{C})$. For any $H\in\mathbb{M}_n(\mathbb{C})$, $H$ is said to be {\it Hermitian} if $H^{\dagger}=H$, and {\it unitary} if $H^{\dagger}H=HH^{\dagger}=I_n$, where $I_n$ denotes the $n$-order {\it identity} matrix. Suppose that $A$ and $B$ are $m$-order and $n$-order complex matrices, respectively. We define the {\it diagonal sum} of $A$ and $B$ to be
\begin{eqnarray*}
   A\oplus B &\triangleq& \left(
                   \begin{array}{cc}
                     A & 0 \\
                     0 & B \\
                   \end{array}
                 \right).
\end{eqnarray*}
Therefore, $A\oplus B$ is an $(m+n)$-order complex matrix.

\par
Let $A=(a_{ij})$ be an $n\times n$ matrix over $\mathbb{C}$. Let $\mathrm{Tr}(A)$ denote the {\it trace} of $A$, i.e., $\mathrm{Tr}(A)=\sum\limits_{i=1}^n a_{ii}$. It is well known that
\begin{eqnarray*}
     \mathrm{Tr}(AB)=\mathrm{Tr}(BA),\quad\mbox{and}\quad\mathrm{Tr}(\lambda_1 A+\lambda_2 B)=\lambda_1\mathrm{Tr}(A)+\lambda_2\mathrm{Tr}(B)
\end{eqnarray*}
where $\lambda_i\in\mathbb{C}$.

\par
Let $V$ be a finite-dimensional vector space over $\mathbb{C}$, and let $\mathcal{B}=\{\eta_1,\eta_2,\cdots,\eta_n\}$ be a basis for $V$ over $\mathbb{C}$. This means that for any vector $\alpha\in V$, there is a unique expression
\begin{eqnarray*}
  \alpha&=&c_1\eta_1+c_2\eta_2+\cdots +c_n\eta_n
\end{eqnarray*}
where $c_i\in\mathbb{C}$. The dimension of $V$, denoted by ${\rm dim} V$, is defined as the cardinality of $\mathcal{B}$. {\color{blue}Let ${\rm span}(\mathcal{B})$ denote the vector space generated by the vectors in $\mathcal{B}$.} Then $V = {\rm span}(\mathcal{B})$. Furthermore, $\mathbb{M}_n(\mathbb{C})$ is a vector space over $\mathbb{C}$ of dimension $n^2$.

\subsection{Some notation on Quantum mechanics}
\label{ssect.2.2}

In quantum theory, to any isolated physical system there is associated a (finite-dimensional) {\em Hilbert space}, denoted $\mathcal{H}$, which is called the state space of the system. In {\em Dirac} notation, the row vector (resp. column vector) $\varphi$ is denoted $\langle\varphi|$ (resp.~ $|\varphi\rangle$). Incidentally, $\langle\varphi|$ is the conjugate transpose of $|\varphi\rangle$, i.e., $\langle\varphi|=|\varphi\rangle^{\dagger}$. The inner product of two vectors $|\varphi\rangle$ and $|\psi\rangle$ is denoted $\langle\varphi|\psi\rangle$. The norm (or length) of the vector $|\varphi\rangle$, denoted $\|\,|\varphi\rangle\,\|$, is defined by $\|\,|\varphi\rangle\,\|=\sqrt{\langle\varphi|\varphi\rangle}$. A vector $|\varphi\rangle$ (resp.~$\langle\varphi|$) is said to be a unit vector if $\|\,|\varphi\rangle\,\|=1$ (resp. $\|\,\langle\varphi|\,\|=1$).

Suppose that $Q=\{q_1,q_2,\cdots,q_m\}$ is the set of basis states of a quantum system. The corresponding Hilbert space is then $\mathcal{H}_m=\mathrm{span}\{|q_i\rangle\,|\,q_i\in Q,\,1\leq i\leq m\}$, where $|q_i\rangle$ is the $m$-dimensional column vector with a $1$ in the $i$-th position and $0$s elsewhere. This space is equipped with the inner product $\langle\cdot|\cdot\rangle$ defined by $\langle\alpha|\beta\rangle=\sum_{i=1}^m x_i^* y_i$, where $\lambda^*$ denotes the complex conjugate of $\lambda\in\mathbb{C}$, and $|\alpha\rangle=(x_1,\dots,x_m)^T$, $|\beta\rangle=(y_1,\dots,y_m)^T$. At any time, the state of the system is a {\em superposition} of the basis states $|q_i\rangle$ ($1\leq i\leq m$) and can be represented by a unit vector $|\phi\rangle=\sum_{i=1}^m c_i|q_i\rangle$ with $c_i\in\mathbb{C}$ satisfying $\sum_{i=1}^m|c_i|^2=1$. One can perform a {\em measurement} on $\mathcal{H}_m$ to extract information about the system. A measurement is described by an observable, i.e., a Hermitian matrix $\mathcal{O}=\sum_{i=1}^s\lambda_i P_i$, where $\lambda_i$ are the eigenvalues and $P_i$ is the projector onto the corresponding eigenspace.

The above mathematical framework describes quantum systems in terms of ``pure states." To handle ``mixed states," the states of a quantum system are represented by a density operator $\rho\in\mathcal{L}(\mathcal{H})$, i.e., an operator that is self-adjoint, positive semi-definite ($\rho\geq 0$), and satisfies ${\rm Tr}(\rho)=1$. The evolution of a closed quantum system is given by a unitary operator $U$ via the map $\rho\mapsto U\rho U^{\dagger}$. More generally, a quantum operation $\mathcal{U}:\mathcal{L}(\mathcal{H}_1)\to\mathcal{L}(\mathcal{H}_2)$ is a trace-preserving completely positive map \cite{1,2,3} of the form $\mathcal{U}(\rho)=\sum_i M_i\rho M_i^{\dagger}$, where $\{M_i\}$ are the Kraus operators satisfying the completeness relation $\sum_i M_i^{\dagger}M_i=I_{\dim\mathcal{H}_1}$. Let $\mathcal{H}=P_1\oplus P_2\oplus\cdots\oplus P_k$ be an orthogonal decomposition. Then, for any $\rho\in\mathcal{L}(\mathcal{H})$, ${\rm Tr}(P_j\rho)$ (equivalently ${\rm Tr}(P_j\rho P_j^{\dagger}$)) gives the probability that the property associated with $P_j$ is observed.

\subsection{On relevant definitions of MM-1QFAs}

For any finite set $S$, $|S|$ denotes the cardinality of $S$. Throughout this paper, $\Sigma$ denotes a non-empty finite alphabet. A {\em word} over the alphabet $\Sigma$ is a finite sequence of symbols chosen from $\Sigma$. Let $\Sigma^*$ denote the set of all words over $\Sigma$. For any word $\omega\in\Sigma^*$, $|\omega|$ denotes the length of $\omega$. Let $\Sigma^n$ denote the set of all words of length $n$ over $\Sigma$, where $n$ is a non-negative integer. Then $\Sigma^*$ can be expressed as
$$\Sigma^* = \{\epsilon\} \cup \Sigma \cup \Sigma^2 \cup \cdots,$$ where $\epsilon$ denotes the empty word.

For a fixed alphabet $\Sigma$, let $M(x)$, where $x\in\Sigma$, be a complex square matrix associated with the symbol $x$. For convenience, we define the formal product $\prod_{i=n}^{1}M(x_i)$ by
\begin{eqnarray*}
\prod\limits_{i=n}^1\,M(x_i)&\triangleq& M(x_n)M(x_{n-1})\cdots M(x_1).
\end{eqnarray*}

\par
Now, we state the definition of an MM-1QFA as follows.
\begin{df}
Formally, an MM-1QFA with $m$ {\color{blue}states} over the alphabet $\Sigma$ is a quadruple $$\mathcal{A}=(Q,\{U(\sigma)\}_{\sigma\in\Sigma\cup\{\$\}}, |\pi\rangle, \mathcal{O})$$ where $Q=\{q_1,q_2,\dots,q_m\}$ is the set of basis states, $|\pi\rangle$ is the initial state vector satisfying $\||\pi\rangle\|=1$, $\$\notin\Sigma$ is the end-marker, and for each $\sigma\in\Sigma\cup\{\$\}$, $U(\sigma)\in\mathbb{M}_m(\mathbb{C})$ is a unitary matrix. The observable $\mathcal{O}$ has possible outcomes in $\{a,r,g\}$ and is completely described by the projectors $P(a)$, $P(r)$, and $P(g)$.
\label{df1}
\end{df}

 The projectors $P(a)$, $P(g)$ and $P(r)$ are given by
$$
  P(a)=\sum_{q\in Q_{acc}}|q\rangle\langle q|,\qquad P(g)=\sum_{q\in Q_{non}}|q\rangle\langle q|, \qquad P(r)=\sum\limits_{q\in Q_{rej}}|q\rangle\langle q|
$$
where $Q_{non}=Q\setminus (Q_{acc}\cup Q_{rej})$ is the set of {\it non-halting} states, $Q_{acc}\subseteq Q$ and $Q_{rej}\subseteq Q$ (with $Q_{acc}\cap Q_{rej}=\emptyset$) are the sets of accepting states and rejecting states, respectively. {\color{blue}Here,  $|q\rangle\langle q|$ denotes the outer product of column vector $|q\rangle$ and row vector $\langle q|$}.

\par
Fed with the input string $x_1x_2\cdots x_n\$$ where $x_1x_2\cdots x_n\in\Sigma^*$, the automaton $\mathcal{A}$ operates as follows. Starting from the initial state $|\pi\rangle$, the unitary operator $U(x_1)$ is applied and a measurement with respect to the observable $\mathcal{O}$ is performed, resulting in a new current state. If the measurement outcome is `$g$', then $U(x_2)$ is applied and $\mathcal{O}$ is measured again. This process continues as long as the measurement outcome is `$g$'. If the measurement outcome is `$a$', the computation halts and the input word is accepted. If the measurement outcome is `$r$', the computation halts and the input word is rejected. Therefore, $\mathcal{A}$ induces a word function $p_{\mathcal{A}}: \Sigma^*\$\rightarrow [0,1]$ in a cumulation manner, i.e.,
\begin{eqnarray}
\label{eq1}
   p_{\mathcal{A}}(x_1x_2\cdots x_n\$)&=&\sum\limits_{k=1}^{n+1}\left\|P(a)U(x_k)\left(\prod\limits_{i=k-1}^{1}
\big(P(g)U(x_i)\big)\right)\,|\pi\rangle\right\|^2
\end{eqnarray}
where $x_{n+1}$ denotes $\$$. By $\prod\limits_{i=0}^1\big(P(g)U(x_i)\big)$ we mean that
$$
   \prod\limits_{i=0}^1\big(P(g)U(x_i)\big)=I_m
$$
i.e., the $m$-order ($m=|Q|$) identity matrix. Further, the probability that $\mathcal{A}$ accepts the word $x_1x_2\cdots x_n$ is defined as
\begin{eqnarray}
\label{eq2}
  \mathcal{P}_{\mathcal{A}}(x_1x_2\cdots x_n)&=&p_{\mathcal{A}}(x_1x_2\cdots x_n\$).
\end{eqnarray}

\begin{df}
\label{df2}
Two MM-1QFAs $\mathcal{A}_1$ and $\mathcal{A}_2$ over $\Sigma$ are said to be equivalent (resp.~$t$-equivalent) if $\mathcal{P}_{\mathcal{A}_1}(\omega)=\mathcal{P}_{\mathcal{A}_2}(\omega)$ for all $\omega\in\Sigma^{*}$ (resp.~for all $\omega\in\Sigma^{*}$ with $|\omega|\leq t$).
\end{df}

\par
{\color{blue}The probability $\cal{P}_{\mathcal{A}}(\omega)$ that $\cal{A}$ accepts the word $\omega$, as given in Eq.~(\ref{eq2}), is somewhat complicated. We now define another ``probability function" of $\mathcal{A}$ accepting the word $\omega$ as follows.
\begin{eqnarray}
\label{eq3}
 \mathcal{F}_{\mathcal{A}}(\omega)&=&
\left\{
  \begin{array}{ll}
    \mathcal{P}_{\mathcal{A}}(x_1x_2\cdots x_n)-\mathcal{P}_{\mathcal{A}}(x_1x_2\cdots x_{n-1}), & \hbox{$\omega=x_1x_2\cdots x_n$;} \\
    \mathcal{P}_{\mathcal{A}}(\epsilon), & \hbox{$\omega=\epsilon$.}
  \end{array}
\right.
\end{eqnarray}
}

\begin{rmk}\label{rmk1}
 Note that if $n=1$ in Eq.~(\ref{eq3}), then $x_1x_2\cdots x_0$ denotes the empty word $\epsilon$. More specifically, we define $\mathcal{F}_{\mathcal{A}}(x)$ to be $\mathcal{P}_{\mathcal{A}}(x)-\mathcal{P}_{\mathcal{A}}(\epsilon)$ for any $x\in\Sigma$.
\end{rmk}

For readability, we introduce the concept of ``$\beta$-equivalence" for MM-1QFAs in terms of Eq.~(\ref{eq3}) as follows.
\begin{df}
\label{df3}
Two MM-1QFAs $\mathcal{A}_1$ and $\mathcal{A}_2$ over the same input alphabet $\Sigma$ are said to be $\beta$-equivalent (resp.~$t$-$\beta$-equivalent) if $\mathcal{F}_{\mathcal{A}_1}(\omega)=\mathcal{F}_{\mathcal{A}_2}(\omega)$ for all $\omega\in\Sigma^*$ (resp.~for all $\omega\in\Sigma^*$ with $|\omega|\leq t$).
\end{df}

\par
The following theorem is the basis that allows us to present a much simpler approach to the equivalence problem for MM-1QFAs.

\begin{thm}
\label{thm4}
 Let $\mathcal{A}_1$ and $\mathcal{A}_2$ be two MM-1QFAs over $\Sigma$. Then $\mathcal{A}_1$ and $\mathcal{A}_2$ are equivalent if and only if they are $\beta$-equivalent.
\end{thm}

\begin{pf}
  We first prove the ``only if" part. Assume that $\mathcal{A}_1$ and $\mathcal{A}_2$ are equivalent. Then
  \begin{eqnarray}\label{eq4}
    \mathcal{P}_{\mathcal{A}_1}(\omega)&=&\mathcal{P}_{\mathcal{A}_2}(\omega)\qquad
    \mbox{($\forall
    \,\omega\in\Sigma^*$)}.
  \end{eqnarray}
  We claim that $\mathcal{F}_{\mathcal{A}_1}(\omega)=\mathcal{F}_{\mathcal{A}_2}(\omega)$ for all $\omega\in\Sigma^*$. By Eqs.~(\ref{eq3}) and (\ref{eq4}), the claim is obvious when $\omega=\epsilon$. For $\omega=x_1x_2\cdots x_n$ with $n\geq 1$, Eq.~(\ref{eq4}) implies
 $$
  \mathcal{P}_{\mathcal{A}_1}(x_1\cdots x_n)-\mathcal{P}_{\mathcal{A}_1}(x_1\cdots x_{n-1})=\mathcal{P}_{\mathcal{A}_2}(x_1\cdots x_n)-\mathcal{P}_{\mathcal{A}_2}(x_1\cdots x_{n-1})
 $$
 i.e., $\mathcal{F}_{\mathcal{A}_1}(x_1\cdots x_n)=\mathcal{F}_{\mathcal{A}_2}(x_1\cdots x_n)$. Thus, the claim holds for all $\omega\in\Sigma^*$.

 \par
 We next prove the ``if" part of Theorem \ref{thm4}. By assumption,
 \begin{eqnarray}
 \label{eq5}
   \mathcal{F}_{\mathcal{A}_1}(\omega)&=&\mathcal{F}_{\mathcal{A}_2}(\omega)\qquad
   \mbox{
   ($\forall\,\omega\in\Sigma^*$)}
 \end{eqnarray}
 It is clear that $\mathcal{P}_{\mathcal{A}_1}(\omega)=\mathcal{P}_{\mathcal{A}_2}(\omega)$ when $\omega=\epsilon$. Assume that $\omega=x_1x_2\cdots x_n$ with $n\geq 1$. For simplicity, denote

   $$ a_n=\mathcal{P}_{\mathcal{A}_1}(x_1\cdots x_n)$$

 and
$$
   b_n=\mathcal{P}_{\mathcal{A}_2}(x_1\cdots x_n)
$$
 for all $n\geq 1$. Setting $a_0=\mathcal{P}_{\mathcal{A}_1}(\epsilon)$ and $b_0=\mathcal{P}_{\mathcal{A}_2}(\epsilon)$, by Eq.~(\ref{eq3}), we have
 $$
   \mathcal{F}_{\mathcal{A}_1}(x_1\cdots x_n)=a_n-a_{n-1}\qquad\mbox{and}\quad \mathcal{F}_{\mathcal{A}_2}(x_1\cdots x_n)=b_n-b_{n-1}.
$$
 Thus,
\begin{eqnarray*}
    \mathcal{P}_{\mathcal{A}_1}(x_1\cdots x_n)&=&a_0+\sum_{k=1}^n(a_k-a_{k-1})\\
    &=&\mathcal{F}_{\mathcal{A}_1}(\epsilon)+\sum_{k=1}^n\mathcal{F}_{\mathcal{A}_1}
    (x_1\cdots x_k)\\
    &=&\mathcal{F}_{\mathcal{A}_2}(\epsilon)+\sum_{k=1}^n\mathcal{F}_{\mathcal{A}_2}
    (x_1\cdots x_k)\qquad\mbox{(by Eq.~(\ref{eq5}))}\\
    &=&b_0+\sum_{k=1}^n(b_k-b_{k-1})=\mathcal{P}_{\mathcal{A}_2}(x_1\cdots x_n).
\end{eqnarray*}
 This completes the proof of Theorem \ref{thm4}.\Q.E.D
\end{pf}

\begin{rmk}
\label{rmk2}
 In fact, it is clear that the proof of Theorem \ref{thm4} can be extended to two that two MM-1QFAs, $\mathcal{A}_1$ and $\mathcal{A}_2$, are $t$-equivalent if and only if they are $t$-$\beta$-equivalent.
\end{rmk}

\par
For convenience, we expand Eq.~(\ref{eq3}) as follows. Note that if $\omega=x_1x_2\cdots x_n$, then
\begin{eqnarray*}
\mathcal{F}_{\mathcal{A}}(\omega)&=&\mathcal{P}_{\mathcal{A}}(x_1x_2\cdots x_n)-\mathcal{P}_{\mathcal{A}}(x_1x_2\cdots x_{n-1})\\
&=&\langle\pi|\,\left(\prod\limits_{i=n-1}^1(P(g)U(x_i))\right)^{\dagger}
U(x_n)^{\dagger}P(a)^{\dagger}P(a)U(x_n)\left(\prod\limits_{i=n-1}^1
(P(g)U(x_i))\right)\,|\pi\rangle\\
&\,\,\,&+\,\,\langle\pi|\,\left(\prod\limits_{i=n}^1(P(g)U(x_i))\right)^{\dagger}
U(\$)^{\dagger}P(a)^{\dagger}P(a)U(\$)\left(\prod\limits_{i=n}^1(P(g)U(x_i))\right)\,
|\pi\rangle\\
&\,\,\,&-\,\,\langle\pi|\,\left(\prod\limits_{i=n-1}^1(P(g)U(x_i))\right)^{\dagger}
U(\$)^{\dagger}
P(a)^{\dagger}P(a)U(\$)\left(\prod\limits_{i=n-1}^1(P(g)U(x_i))\right)\,|\pi\rangle.
\end{eqnarray*}

\par
Setting $A(\sigma)=P(g)U(\sigma)$ for each $\sigma\in\Sigma$, and noting that $P(a)^2=P(a)$ and $P(a)^{\dagger}=P(a)$, we obtain
\begin{eqnarray}
\label{eq6}
  \mathcal{F}_{\mathcal{A}}(\omega)&=&\langle\pi|\,\eta_{\mathcal{A}}(\omega)\,|
  \pi\rangle,
\end{eqnarray}
where
\begin{eqnarray*}
   \eta_{\mathcal{A}}(\omega)&=&\left\{
                                            \begin{array}{lll}
                                              \Big(\prod\limits_{i=n-1}^1
                                              A(x_i)\Big)^{\dagger}
   \delta_{\mathcal{A}}(x_n)\Big(\prod\limits_{i=n-1}^1A(x_i)\Big), & \hbox{$\omega=x_1x_2\cdots x_n\in\Sigma^n$;} \\
\\
                                              U(\$)^{\dagger}P(a)U(\$), & \hbox{$\omega=\epsilon$.}
                                            \end{array}
                                          \right.
\end{eqnarray*}
and $\delta_{\mathcal{A}}(x_n)$ is given by
$$
  \delta_{\mathcal{A}}(x_n)=U(x_n)^{\dagger}P(a)U(x_n)+A(x_n)^{\dagger}
U(\$)^{\dagger}P(a)U(\$)A(x_n)-U(\$)^{\dagger}P(a)U(\$).
$$

\par
We further introduce the following auxiliary definitions, which will be needed in the sequel.

\begin{df}\label{df4}
  Let $\mathcal{A}_i=(Q_i,\{U_i(\sigma)\}_{\sigma\in\Sigma\cup\{\$\}},|\pi_i\rangle,
  \mathcal{O}_i)$, $i=1,2$, be two MM-1QFAs over the alphabet $\Sigma$, where $\mathcal{O}_1=\{P_1(a),P_1(g),P_1(r)\}$ and $\mathcal{O}_2=\{P_2(a),P_2(g),P_2(r)\}$. The {\em diagonal sum} of $\mathcal{A}_1$ and $\mathcal{A}_2$, denoted by $\mathcal{A}_1\oplus\mathcal{A}_2$, is an MM-1QFA defined as
  $$
    \mathcal{A}=\mathcal{A}_1\oplus\mathcal{A}_2=(Q,\{U(\sigma)\}_{\sigma\in\Sigma\cup
    \{\$\}},|\vartheta\rangle,\mathcal{O}),
  $$
  where $Q=Q_1\cup Q_2$ with $Q_1\cap Q_2=\emptyset$, $U(\sigma)=U_1(\sigma)\oplus U_2(\sigma)$ for each $\sigma\in\Sigma\cup\{\$\}$, $|\vartheta\rangle\in\mathcal{H}_{|Q_1|+|Q_2|}$ is an arbitrary unit vector, and $\mathcal{O}=\{P_1(a)\oplus P_2(a),P_1(g)\oplus P_2(g), P_1(r)\oplus P_2(r)\}$.
\end{df}

\par
It should be noted that the initial vector $|\vartheta\rangle$ of $\mathcal{A}$ is arbitrary. Of particular importance are the following two vectors:
\begin{eqnarray}\label{eq7}
  |\varphi\rangle\,=\,\left(
                     \begin{array}{c}
                       |\pi_1\rangle \\
                       0 \\
                     \end{array}
                   \right),
\qquad |\psi\rangle\,=\,\left(
                          \begin{array}{c}
                            0 \\
                            |\pi_2\rangle \\
                          \end{array}
                        \right).
\end{eqnarray}

\par
With respect to the above vectors, we introduce the following technical definition.
\begin{df}\label{df5}
  Let $\mathcal{A}_i=(Q_i,\{U_i(\sigma)\}_{\sigma\in\Sigma\cup\{\$\}},|\pi_i\rangle,
  \mathcal{O}_i)$, $i=1,2$, be two MM-1QFAs over $\Sigma$. Let $\mathcal{A}=\mathcal{A}_1\oplus\mathcal{A}_2$. Then the vectors $|\varphi\rangle$ and $|\psi\rangle$, defined in Eqs.~(\ref{eq7}), are said to be equivalent with respect to $\mathcal{A}$ (resp.~$t$-equivalent with respect to $\mathcal{A}$), if
  \begin{eqnarray}\label{eq8}
  \langle\varphi|\, \eta_{\mathcal{A}}(\omega)\,|\varphi\rangle&=&\langle\psi|\, \eta_{\mathcal{A}}(\omega)|\psi\rangle
\end{eqnarray}
for all $\omega\in\Sigma^{*}$ (resp.~for all $\omega\in\Sigma^{*}$ with $|\omega|\leq t$).
\end{df}

\begin{rmk}\label{rmk3}
In fact, the left-hand side of Eq.~(\ref{eq8}) is $\mathcal{F}_{\mathcal{A}_1}(\omega)$, and the right-hand side  is $\mathcal{F}_{\mathcal{A}_2}(\omega)$. To see this, it is straightforward to verify that
\begin{eqnarray}\label{eq9}
  \eta_{\mathcal{A}}(\omega)&=&\left(
                                           \begin{array}{cc}
                                             \eta_{\mathcal{A}_1}(\omega) & 0 \\
                                             0 & \eta_{\mathcal{A}_2}(\omega) \\
                                           \end{array}
                                         \right)
\end{eqnarray}
for all $\omega\in\Sigma^{*}$.
Hence, 
\begin{eqnarray*}
 \langle\varphi|\,\eta_{\mathcal{A}}(\omega)\,|\varphi\rangle=\langle\pi_1|\,
 \eta_{\mathcal{A}_1}(\omega)\,|\pi_1\rangle=\mathcal{F}_{\mathcal{A}_1}(\omega)
\end{eqnarray*}
and
\begin{eqnarray*}
  \langle\psi|\,\eta_{\mathcal{A}}(\omega)\,|\psi\rangle=\langle\pi_2|\,
  \eta_{\mathcal{A}_2}(\omega)\,|\pi_2\rangle=\mathcal{F}_{\mathcal{A}_2}(\omega).
\end{eqnarray*}
\end{rmk}

\par
Let $\mathcal{A}=(Q,\{U(\sigma)\}_{\sigma\in\Sigma\cup\{\$\}},|\pi\rangle,\mathcal{O})$ be an MM-1QFA. Suppose that $\omega=x_1x_2\cdots x_n\in\Sigma^{*}$ and $y\in\Sigma$ are arbitrary. Then
\begin{eqnarray}\label{eq10}
  \eta_{\mathcal{A}}(y\omega)&=&\left[\left(\prod\limits_{i=n-1}^1A(x_i)\right)A(y)
  \right]^{\dagger}
  \delta_{\mathcal{A}}(x_n)\left[\left(\prod\limits_{i=n-1}^1A(x_i)\right)A(y)\right]
  \nonumber\\
  &=&A(y)^{\dagger}\left[\left(\prod\limits_{i=n-1}^1A(x_i)\right)^{\dagger}
  \delta_{\mathcal{A}}(x_n)\left(\prod\limits_{i=n-1}^1A(x_i)\right)\right]A(y)
  \nonumber\\
  &=&A(y)^{\dagger}\,\eta_{\mathcal{A}}(\omega)\,A(y)
\end{eqnarray}

\begin{rmk}
 Eq.~(\ref{eq10}) pays a key role in the proof of Lemma \ref{lem6} and is inspired by the proof of Lemma {\bf 8} in \cite{27} (Li and Qiu) and by the proof of Theorem {\bf 1} in \cite{4}(Carlyle).
\end{rmk}

\subsection{On relevant definitions of E-1QFAs}

As mentioned earlier, an E-1QFA is a theoretical model for a quantum computer with finite workspace \cite{22} and can be viewed as a generalization of MM-1QFA. In what follows, we first present the definition of an E-1QFA.
\begin{df}[{{\it Modification of \cite{22}}}]\label{df6}
An E-1QFA over the alphabet $\Sigma$ is a sextuple
\begin{eqnarray*}
\mathcal{A}&=&(Q,Q_{acc},Q_{rej},\{\mathcal{U}_{\sigma}\}_{\sigma\in\Sigma\cup
\{\#,\$\}},\rho,\mathcal{O})
\end{eqnarray*}
where $Q$ is a finite set of states,  $Q_{acc}\subseteq Q$ and $Q_{rej}\subseteq Q$ are the accepting and rejecting sets of states, respectively. For each symbol $\sigma\in\Sigma\cup\{\#,\$\}$ where $\#$ and $\$$ are the left and right end-marker,respectively, $\mathcal{A}$ has a corresponding ``superoperator" $\mathcal{U}_{\sigma}$;\footnote{Here, the ``superoperator" \cite{22} is realized by a composition of a finite sequence of unitary transformations and orthogonal measurements on the space $\mathbb{C}^Q$ (i.e., $\mathcal{H}_Q$; see Subsection \ref{ssect.2.2}). However, if arbitrary POVM measurements are allowed instead of orthogonal measurements, then the set of ``superoperators" consists of all possible quantum operations \cite{33}.} The initial state of  $\mathcal{A}$ is the density matrix $\rho=|q_0\rangle\langle q_0|$ with $q_0\in Q$, and $\mathcal{O}=\{P_a,P_g,P_r\}$, where $P_a$, $P_g$, and $P_r$ are the orthogonal projection onto span$\{|q\rangle|q\in Q_{acc}\}$, span$\{|q\rangle|q\in Q\backslash(Q_{acc}\cup Q_{rej})\}$, and span$\{|q\rangle|q\in Q_{rej}\}$, respectively.
\end{df}

\par
The computing procedure of an E-1QFA is similar to that of an MM-1QFA. For further details, we refer to \cite{22} (cf. \cite{22}, Section 3.2). Thus, for a word $\omega=x_1x_2\cdots x_n\in\Sigma^*$, an E-1QFA $\mathcal{A}$ induces the word function
\begin{eqnarray}\label{eq11}
  p_{\mathcal{A}}(\#\omega\$)&=&\mathrm{Tr}\left(\sum_{k=0}^{n+1}(P_a\circ
  \mathcal{U}_{x_k})
  \circ\left[\prod_{i=k-1}^0
  \left(P_g\circ\mathcal{U}_{x_i}\right)\right](\rho)\right),
\end{eqnarray}
where $x_0=`\#$' and $x_{n+1}=`\$$'.
The acceptance probability of $\mathcal{A}$ on $\omega$ is then defined by
\begin{eqnarray}\label{eq12}
  \mathcal{P}_{\mathcal{A}}(\omega)&=&p_{\mathcal{A}}(\#\omega\$).
\end{eqnarray}
In Eq.~(\ref{eq11}), the formal product $\prod\limits_{i=m}^0\mathcal{U}_i$ is defined as
\begin{eqnarray*}
 \prod_{i=m}^0\mathcal{U}_i&=&\mathcal{U}_m\circ\mathcal{U}_{m-1}
 \circ\cdots\circ\mathcal{U}_0.
\end{eqnarray*}
By convention, $\prod\limits_{i=-1}^0(P_g\circ U_{x_i})$ denotes the identity superoperator $\mathcal{I}$ on $\mathcal{L}(\mathcal{H}_Q)$. The composition $P_g\circ\mathcal{U}$ is defined by
\begin{eqnarray*}
P_g\circ\mathcal{U}(\rho')&=&P_g\left(\sum_iM_i\rho'M_i^{\dagger}\right)P_g^{\dagger}\\
&=&\sum_i\Big[(P_gM_i)\rho'(P_gM_i)^{\dagger}\Big]
\end{eqnarray*}
for any $\rho'\in\mathcal{L}(\mathcal{H}_Q)$, where $\{M_i\}$ are the Kraus operators of $\mathcal{U}$. The operator $P_a\circ\mathcal{U}$ is defined analogously.

\begin{df}\label{df7}
 Two E-1QFAs $\mathcal{A}_1$ and $\mathcal{A}_2$ over the same alphabet $\Sigma$ are said to be equivalent (resp.~$t$-equivalent), if $\mathcal{P}_{\mathcal{A}_1}(\omega)=\mathcal{P}_{\mathcal{A}_2}(\omega)$ for all $\omega\in\Sigma^*$ (resp.~for all $\omega\in\Sigma^*$ with $|\omega|\leq t$).
\end{df}

\par
Similarly, the acceptance probability $\mathcal{P}_{\mathcal{A}}(\omega)$  given by Eq.~(\ref{eq12}) is defined in a ``cumulation" manner. We can also define  a ``non-cumulative" manner as follows:
\begin{eqnarray}\label{eq13}
  \mathcal{F}_{\mathcal{A}}(\omega)&=&\left\{
                                        \begin{array}{ll}
                                          \mathcal{P}_{\mathcal{A}}(x_1x_2\cdots x_n)-\mathcal{P}_{\mathcal{A}}(x_1x_2\cdots x_{n-1}), & \hbox{$\omega=x_1x_2\cdots x_n$;} \\
                                          \mathcal{P}_{\mathcal{A}}(\epsilon), & \hbox{$\omega=\epsilon$.}
                                        \end{array}
                                      \right.
\end{eqnarray}

\par
Analogously to the MM-1QFAs, we define  ``$\beta$-equivalence" for E-1QFAs using Eq.~(\ref{eq13}).
\begin{df}\label{df8}
Two E-1QFAs $\mathcal{A}_1$ and $\mathcal{A}_2$ over the same alphabet $\Sigma$ are said to be $\beta$-equivalent (resp.~$t$-$\beta$-equivalent) if $\mathcal{F}_{\mathcal{A}_1}(\omega)=\mathcal{F}_{\mathcal{A}_2}(\omega)$ for all $\omega\in\Sigma^*$ (resp.~for all $\omega\in\Sigma^*$ with $|\omega|\leq t$).
\end{df}

\par
The following theorem allows us to reduce the equivalence problem for E-1QFAs to that for MM-1QFAs.

\begin{thm}\label{thm5}
  Let $\mathcal{A}_1$ and $\mathcal{A}_2$ be two E-1QFAs over the same alphabet $\Sigma$. Then $\mathcal{A}_1$ and $\mathcal{A}_2$ are equivalent (resp.~$t$-equivalent) if and only if they are $\beta$-equivalent (resp.~$t$-$\beta$-equivalent).
\end{thm}

\begin{pf}
 The proof is similar to that of Theorem \ref{thm4}, and the details are omitted. \Q.E.D
\end{pf}

\par
Note that if $\omega=x_1x_2\cdots x_n$ with $n\geq 1$, then $\mathcal{F}_{\mathcal{A}}(\omega)$ can be expressed as
\begin{eqnarray}\label{eq14}
\mathcal{F}_{\mathcal{A}}(\omega)=\mathrm{Tr}\left(\Big(P_a\circ \mathcal{U}_{x_n}+(P_a\circ\mathcal{U}_{\$})\circ (P_g\circ\mathcal{U}_{x_n})-P_a\circ\mathcal{U}_{\$}\Big)\circ\prod_{i=n-1}^0
\Big(P_g\circ\mathcal{U}_{x_i}\Big)(\rho)\right)
\end{eqnarray}

\par
We can rewrite Eq.~(\ref{eq14}) as
\begin{eqnarray*}
\mathcal{F}_{\mathcal{A}}(\omega)&=&\mathrm{Tr}\left((P_a\circ\mathcal{U}_{x_n}+
(P_a\circ\mathcal{U}_{\$})\circ (P_g\circ\mathcal{U}_{x_n})-P_a\circ\mathcal{U}_{\$})(\rho')\right),
\end{eqnarray*}
where
\begin{eqnarray*}
\rho'&=&\prod_{i=n-1}^0\Big(P_g\circ\mathcal{U}_{x_i}\Big)(\rho)\nonumber\\
&=&\sum_{i_{x_{n-1}}}(P_gM_{i_{x_{n-1}}})\left[\cdots\left[\sum_{i_{x_0}}
(P_gM_{i_{x_0}})|q_0\rangle\langle q_0|(P_gM_{i_{x_0}})^{\dagger}\right]\cdots\right](P_gM_{i_{x_{n-1}}})^{\dagger}\\
&=&\sum_{i_{x_{n-1}}}\cdots\sum_{i_{x_0}}\left[(P_gM_{i_{x_{n-1}}})\cdots
(P_gM_{i_{x_0}})|q_0\rangle\langle q_0|(P_gM_{i_{x_0}})^{\dagger}\cdots(P_gM_{i_{x_{n-1}}})^{\dagger}\right].
\end{eqnarray*}

Setting $P_aM_j=A_j$ and $P_gM_j=B_j$ for every Kraus operator $M_j$, a direct calculation yields
\begin{eqnarray*}
  \mathrm{Tr}(P_a\circ\mathcal{U}_{x_n}(\rho'))&=&
  \mathrm{Tr}\left(\sum_{i_{x_n}}\sum_{i_{x_{n-1}}}\cdots\sum_{i_{x_0}}A_{i_{x_n}}
  B_{i_{x_{n-1}}}\cdots B_{i_{x_0}}|q_0\rangle\langle q_0|B_{i_{x_0}}^{\dagger}\cdots B_{i_{x_{n-1}}}^{\dagger}A_{i_{x_n}}^{\dagger}\right)\\
  &&\qquad\mbox{(by the commutative law of Tr, we have)}\\
  &=&\mathrm{Tr}\left(\langle q_0|\left[\sum_{i_{x_0}}\cdots\sum_{i_{x_{n-1}}}\sum_{i_{x_n}}
  B_{i_{x_0}}^{\dagger}\cdots B_{i_{x_{n-1}}}^{\dagger}A_{i_{x_n}}^{\dagger}A_{i_{x_n}}B_{i_{x_{n-1}}}\cdots B_{i_{x_0}}\right]|q_0\rangle\right)\\
  &=& \langle q_0|\left[\sum_{i_{x_0}}\cdots\sum_{i_{x_{n-1}}}B_{i_{x_0}}^{\dagger}\cdots B_{i_{x_{n-1}}}^{\dagger}\left(\sum_{i_{x_n}}A_{i_{x_n}}^{\dagger}A_{i_{x_n}}\right)
  B_{i_{x_{n-1}}}\cdots B_{i_{x_0}}\right]|q_0\rangle,
\end{eqnarray*}\\
and similarly for other two terms:

$\mathrm{Tr}(P_a\circ\mathcal{U}_{\$}\circ P_g\circ\mathcal{U}_{x_n}(\rho'))\,\,=$\\
\begin{eqnarray*}
\langle q_0|\left[\sum_{i_{x_0}}\cdots\sum_{i_{x_{n-1}}}B_{i_{x_0}}^{\dagger}\cdots B_{i_{x_{n-1}}}^{\dagger}\left(\sum_{i_{x_n}}\sum_{i_{x_{n+1}}}B_{i_{x_n}}^{\dagger}
A_{i_{x_{n+1}}}^{\dagger}A_{i_{x_{n+1}}}B_{i_{x_n}}\right)B_{i_{x_{n-1}}}\cdots B_{i_{x_0}}\right]|q_0\rangle,
\end{eqnarray*}\\

\begin{eqnarray*}
 \mathrm{Tr}(P_a\circ\mathcal{U}_{\$}(\rho'))
 &=&\langle q_0|\left[\sum_{i_{x_0}}\cdots\sum_{i_{x_{n-1}}}B_{i_{x_0}}^{\dagger}\cdots B_{i_{x_{n-1}}}^{\dagger}\left(\sum_{i_{x_{n+1}}}A_{i_{x_{n+1}}}^{\dagger}
 A_{i_{x_{n+1}}}\right)B_{i_{x_{n-1}}}\cdots B_{i_{x_0}}\right]|q_0\rangle.
\end{eqnarray*}

It is then straightforward to verify that
\begin{eqnarray*}
  \mathcal{F}_{\mathcal{A}}(\omega)&=&\mathrm{Tr}\left(P_a\circ\mathcal{U}_{x_n}
  (\rho')\right)+\mathrm{Tr}\left((P_a\circ\mathcal{U}_{\$})\circ (P_g\circ\mathcal{U}_{x_n})(\rho')\right)-\mathrm{Tr}\left(P_a\circ
  \mathcal{U}_{\$}(\rho')\right) \\
  &=&\langle q_0|\left[\sum_{i_{x_0}}\cdots\sum_{i_{x_{n-1}}}B_{i_{x_0}}^{\dagger}\cdots B_{i_{x_{n-1}}}^{\dagger}\xi_{\mathcal{A}}(x_n)\,\,B_{i_{x_{n-1}}}\cdots B_{i_{x_0}}\right]|q_0\rangle,
\end{eqnarray*}
where 
\begin{eqnarray*}
  \xi_{\mathcal{A}}(x_n)&=&\sum_{i_{x_n}}A_{i_{x_n}}^{\dagger}A_{i_{x_n}}+
  \sum_{i_{x_n}}\sum_{i_{x_{n+1}}}B_{i_{x_n}}^{\dagger}A_{i_{x_{n+1}}}^{\dagger}
  A_{i_{x_{n+1}}}B_{i_{x_n}}
  -\sum_{i_{x_{n+1}}}A_{i_{x_{n+1}}}^{\dagger}A_{i_{x_{n+1}}}.
\end{eqnarray*}

\par
Since an E-1QFA has a left end-marker `\#' (unlike the MM-1QFA), the approach used for MM-1QFAs cannot be applied directly. A more careful treatment is required. Thus, we define 
\begin{eqnarray*}
    \vartheta_{\mathcal{A}}(\omega)&=&\sum_{i_{x_1}}\cdots\sum_{i_{x_{n-1}}}
    B_{i_{x_1}}^{\dagger}\cdots B_{i_{x_{n-1}}}^{\dagger}\xi_{\mathcal{A}}(x_n)B_{i_{x_{n-1}}}\cdots B_{i_{x_1}}
\end{eqnarray*}
and
\begin{eqnarray}\label{eq15}
    \theta_{\mathcal{A}}(\omega)&=&\sum_{i_{x_0}}B^{\dagger}_{i_{x_0}}
    \left(\sum_{i_{x_1}}\cdots\sum_{i_{x_{n-1}}}B_{i_{x_1}}^{\dagger}\cdots B_{i_{x_{n-1}}}^{\dagger}\xi_{\mathcal{A}}(x_n)B_{i_{x_{n-1}}}\cdots B_{i_{x_1}}\right) B_{i_{x_0}}\nonumber\\
    &=&\sum_{i_{x_0}}B_{i_{x_0}}^{\dagger}\vartheta_{\mathcal{A}}(\omega)B_{i_{x_0}}
\end{eqnarray}
for any $\omega=x_1x_2\cdots x_n\in\Sigma^*$.

\par
The following technical definition of the ``{\it diagonal sum}" of E-1QFAs pays the same role as the corresponding definition for MM-1QFAs.

\begin{df}\label{df9}
Let $\mathcal{A}_i=(Q_i,Q_{acc,i},Q_{rej,i},\{\mathcal{U}_{\sigma}^{(i)}\}_{\sigma\in\Sigma
\cup\{\#,\$\}},\rho_i,\mathcal{O}_i)$, $i=1,2$, be two E-1QFAs over $\Sigma$, where $\mathcal{O}_i=\{P_a^{(i)},P_g^{(i)},P_r^{(i)}\}$, and $\rho_i=|q_0^{(i)}\rangle\langle q_0^{(i)}|$. The {\em diagonal sum} of $\mathcal{A}_1$ and $\mathcal{A}_2$, denoted $\mathcal{A}_1\oplus\mathcal{A}_2$, is defined as
\begin{eqnarray*}
   \mathcal{A}&\triangleq&\mathcal{A}_1\oplus\mathcal{A}_2=(Q,Q_{acc},Q_{rej},
   \{\mathcal{U}_{\sigma}\}_{\sigma\in\Sigma\cup\{\#,\$\}},\varrho,\mathcal{O}),
\end{eqnarray*}
where $Q=Q_1\cup Q_2$ with $Q_1\cap Q_2=\emptyset$, $\mathcal{U}_{\sigma}=\mathcal{U}_{\sigma}^{(1)}\oplus\mathcal{U}_{\sigma}^{(2)}$
\footnote{Here, if $\mathcal{U}_{\sigma}^{(1)}$ and $\mathcal{U}_{\sigma}^{(2)}$ are given by the operators sets $\{E_i\}$ and $\{Z_j\}$, respectively, then $\mathcal{U}_{\sigma}$ can be defined to be given by the operators set $\{M_i\}\triangleq\{E_i\oplus Z_i\}$. It is not hard to see that $\sum_iM_i^{\dagger}M_i=\left(
                          \begin{array}{cc}
                            \sum_iE_i^{\dagger}E_i & 0 \\
                            0 & \sum_iZ_i^{\dagger}Z_i \\
                          \end{array}
                        \right)$ and $\mathcal{U}_{\sigma}(\rho)=\left(
                                        \begin{array}{cc}
                                          \sum_iE_i\rho_1 E_i^{\dagger} & 0 \\
                                          0 & \sum_iZ_i\rho_2 Z_i^{\dagger} \\
                                        \end{array}
                                      \right)=\left(
                                                \begin{array}{cc}
                                                  \mathcal{U}_{\sigma}^{(1)}(\rho_1) & 0 \\
                                                  0 & \mathcal{U}_{\sigma}^{(2)}(\rho_2) \\
                                                \end{array}
                                              \right)
                                      $ for any $\rho=\rho_1\oplus \rho_2$.
}, $\varrho\in\mathcal{L}(\mathcal{H}_{Q_1\cup Q_2})$ is an arbitrary density matrix, and $\mathcal{O}=\{P_a^{(1)}\oplus P_a^{(2)}, P_g^{(1)}\oplus P_g^{(2)}, P_r^{(1)}\oplus P_r^{(2)}\}$.
\end{df}

\par
As in the MM-1QFA case, the initial state $\varrho$ is arbitrary. Of particular importance are the following two density matrices:
\begin{eqnarray}\label{eq16}
 \varphi=\left(
             \begin{array}{cc}
               \rho_1 & 0 \\
               0 & 0 \\
             \end{array}
           \right),\qquad
   \psi=\left(
          \begin{array}{cc}
            0 & 0 \\
            0 & \rho_2 \\
          \end{array}
        \right).
\end{eqnarray}

\par
We now introduce the following definition.
\begin{df}\label{df10}
  Let $\mathcal{A}_i=(Q_i,Q_{acc,i},Q_{rej,i},\{\mathcal{U}_{\sigma}^{(i)}\}_{\sigma
  \in\Sigma\cup\{\#,\$\}},\rho_i,\mathcal{O}_i)$, $i=1,2$, be two E-1QFAs over $\Sigma$, where $\mathcal{O}_i=\{P_a^{(i)},P_g^{(i)},P_r^{(i)}\}$ and $\rho_i=|q_0^{(i)}\rangle\langle q_0^{(i)}|$. Let $\mathcal{A}=\mathcal{A}_1\oplus\mathcal{A}_2$. Then
   the density matrices $\varphi$ and $\psi$, defined in Eqs.~(\ref{eq16}), are said to be equivalent with respect to $\mathcal{A}$ (resp.~$t$-equivalent with respect to $\mathcal{A}$)
   if
   \begin{eqnarray}\label{eq17}
   (\langle q_0^{(1)}|,{\bf 0})\,\,\theta_{\mathcal{A}}(\omega)\,\,\left(
                                                          \begin{array}{c}
                                                            |q_0^{(1)}\rangle \\
                                                            {\bf 0} \\
                                                          \end{array}
                                                        \right) &=&
      ({\bf 0},\langle q_0^{(2)}|)\,\,\theta_{\mathcal{A}}(\omega)\,\,\left(
                                           \begin{array}{c}
                                                           {\bf 0} \\
                                                           |q_0^{(2)}\rangle \\
                                                         \end{array}
                                                       \right)
      \end{eqnarray}
   for all $\omega\in\Sigma^*$ (resp.~for all $\omega\in\Sigma^*$ with $|\omega|\leq t$).
\end{df}

\begin{rmk}
\label{rmk5}
  It is easy to see that
  \begin{eqnarray}\label{eq18}
   \theta_{\mathcal{A}}(\omega)&=&\left(
                                 \begin{array}{cc}
                                   \theta_{\mathcal{A}_1}(\omega) & 0 \\
                                   0 & \theta_{\mathcal{A}_2}(\omega) \\
                                 \end{array}
                               \right)
  \end{eqnarray}
  for all $\omega\in\Sigma^*$.
  Consequently, 
  \begin{eqnarray*}
    (\langle q_0^{(1)}|,{\bf 0})\,\,\theta_{\mathcal{A}}(\omega)\,\,\left(
                                                                \begin{array}{c}
                                                                  |q_0^{(1)}\rangle \\
                                                                  {\bf 0} \\
                                                                \end{array}
                                                              \right)= \langle q_0^{(1)}|
                                                              \theta_{\mathcal{A}_1}
                                                              (\omega)|q_0^{(1)}
                                                              \rangle=
                                                              \mathcal{F}_{\mathcal{A}_1}(\omega)
\end{eqnarray*}
and
\begin{eqnarray*}
    ({\bf 0},\langle q_0^{(2)}|)\,\,\theta_{\mathcal{A}}(\omega)\,\,\left(
                                                               \begin{array}{c}
                                                                 {\bf 0} \\
                                                                 |q_0^{(2)}\rangle \\
                                                               \end{array}
                                                             \right)=
    \langle q_0^{(2)}|\theta_{\mathcal{A}_2}(\omega)|q_0^{(2)}\rangle=\mathcal{F}_{\mathcal{A}_2}
    (\omega).
\end{eqnarray*}
Namely, the left-hand side of Eq.~(\ref{eq17}) is $\mathcal{F}_{\mathcal{A}_1}(\omega)$, and the right-hand side is $\mathcal{F}_{\mathcal{A}_2}(\omega)$.
\end{rmk}

\par
In the following, we derive a relation similar to Eq.~(\ref{eq10}).

 Let $\mathcal{A}_i=(Q_i,Q_{acc,i},Q_{rej,i},
\{\mathcal{U}^{(i)}_{\sigma}\}_{\sigma\in\Sigma\cup\{\#,\$\}},\rho_i,
\mathcal{O}_i)$, $i=1,2$, be two E-1QFAs, and let $\mathcal{A}=\mathcal{A}_1\oplus\mathcal{A}_2$. Suppose that $\omega=x_1x_2\cdots x_n\in\Sigma^*$ and $y\in\Sigma$ are arbitrary. Then
\begin{eqnarray}\label{eq19}
  \vartheta_{\mathcal{A}}(y\omega)&=&\sum_{i_y}\sum_{i_{x_1}}\cdots\sum_{i_{x_{n-1}}}
  B_{i_y}^{\dagger}B_{i_{x_1}}^{\dagger}\cdots B_{i_{x_{n-1}}}^{\dagger}\xi_{\mathcal{A}}(x_n)B_{i_{x_{n-1}}}\cdots B_{i_{x_1}}B_{i_y}\nonumber\\
  &=&\sum_{i_y}B_{i_y}^{\dagger}\left[\sum_{i_{x_1}}\cdots\sum_{i_{x_{n-1}}}
  B_{i_{x_1}}^{\dagger}\cdots B_{i_{x_{n-1}}}^{\dagger}\xi_{\mathcal{A}}(x_n)B_{i_{x_{n-1}}}\cdots B_{i_{x_1}}\right]B_{i_y}\nonumber\\
  &=&\sum_{i_y}B_{i_y}^{\dagger}\vartheta_{\mathcal{A}}(\omega)B_{i_y}.
\end{eqnarray}

\begin{rmk}
 Just as Eq.~(\ref{eq10}) plays a key role in the proof of Lemma \ref{lem6}, Eq.~(\ref{eq19}) plays a similar role in the proof of Lemma \ref{lem9}.
\end{rmk}

\section{Proof of Theorem \ref{thm1}}
\label{sec:proof_of_theorem1}

In this section, we present our approach to the equivalence problem for MM-1QFAs. Let us first introduce some convenient notation.

\par
For each $i\geq 0$, let $H_{\mathcal{A}}(i)$ denote the set $\{\eta_{\mathcal{A}}(\omega)|\omega\in\Sigma^*,|\omega|\leq i\}$, where $H_{\mathcal{A}}(0)=\{U(\$)^{\dagger}P(a)U(\$)\}$, and let $\mathcal{V}_{\mathcal{A}}(i)$ be the vector space spanned by $H_{\mathcal{A}}(i)$, i.e., $\mathcal{V}_{\mathcal{A}}(i)=$span$\{H_{\mathcal{A}}(i)\}$. Then it is clear that $\mathcal{V}_{\mathcal{A}}(i)\subseteq\mathcal{V}_{\mathcal{A}}(i+1)$ since $H_{\mathcal{A}}(i)\subseteq H_{\mathcal{A}}(i+1)$. We prove the following lemma.

\begin{lem}\label{lem6}
  Let $\mathcal{A}=(Q,\{U(\sigma)\}_{\sigma\in\Sigma\cup\{\$\}},|\pi\rangle,\mathcal{O})$ be an MM-1QFA. Then there exists an integer $l<|Q|^2$, such that $\mathcal{V}_{\mathcal{A}}(l)=\mathcal{V}_{\mathcal{A}}(l+j)$ for all $j\geq 1$.
\end{lem}
\begin{pf}
 We first show that there exists an integer $l<|Q|^2$ such that $\mathcal{V}_{\mathcal{A}}(l)=\mathcal{V}_{\mathcal{A}}(l+1)$. Suppose no such an integer exists. Then for all $i\geq 0$, $\mathcal{V}_{\mathcal{A}}(i)\neq\mathcal{V}_{\mathcal{A}}(i+1)$. This gives the chain
 \begin{eqnarray*}
   \mathcal{V}_{\mathcal{A}}(0)\subset\mathcal{V}_{\mathcal{A}}(1)\subset\cdots\subseteq
   \mathbb{M}_{|Q|}(\mathbb{C}).
 \end{eqnarray*}
 Since dim$\mathbb{M}_{|Q|}(\mathbb{C})=|Q|^2$ and dim$\mathcal{V}_{\mathcal{A}}(0)\geq 1$, we have dim$\mathcal{V}_{\mathcal{A}}(|Q|^2)\geq |Q|^2+1$ which contradicts $\mathcal{V}_{\mathcal{A}}(|Q|^2)\subseteq\mathbb{M}_{|Q|}(\mathbb{C})$.

 \par
 We next show that $\mathcal{V}_{\mathcal{A}}(l)=\mathcal{V}_{\mathcal{A}}(l+j)$ for all $j\geq 1$ by induction on $j$. The case $j=1$ has already been shown. Assume the statement holds  for all $j<m$ where $m>1$, and consider the case $j=m$. Note that $$H_{\mathcal{A}}(l+m)=H_{\mathcal{A}}(l+(m-1))\cup\{\eta_{\mathcal{A}}
 (\omega)|\omega\in\Sigma^*,|\omega|=l+m\}$$ and $\mathcal{V}_{\mathcal{A}}(l+m)=\mathrm{span}\{H_{\mathcal{A}}(l+m)\}$. Thus, for any $\eta\in\mathcal{V}_{\mathcal{A}}(l+m)$, we can write
 \begin{eqnarray*}
   \eta&=&\sum_{i_1}a_{i_1}\eta_{\mathcal{A}}(\omega_{i_1})+\sum_{i_2}a_{i_2}
   \eta_{\mathcal{A}}(\omega_{i_2})
 \end{eqnarray*}
 where $\eta_{\mathcal{A}}(\omega_{i_1})\in H_{\mathcal{A}}(l+(m-1))$ and $\eta_{\mathcal{A}}(\omega_{i_2})\in\{\eta_{\mathcal{A}}(\omega)|\omega\in\Sigma^*,
 |\omega|=l+m\}$. Clearly, $\sum_{i_1}a_{i_1}\eta_{\mathcal{A}}(\omega_{i_1})\in\mathcal{V}_{\mathcal{A}}
 (l+(m-1))$. We claim that $\sum_{i_2}a_{i_2}\eta_{\mathcal{A}}(\omega_{i_2})\in\mathcal{V}_{\mathcal{A}}
 (l+(m-1))$ as well.
 To see this, it suffices to prove that each $\eta_{\mathcal{A}}(\omega_{i_2})\in\{\eta_{\mathcal{A}}(\omega)|\omega\in\Sigma^*,
 |\omega|=l+m\}$ can be expressed as $\eta_{\mathcal{A}}(\omega_{i_2})=\sum_zb_z\eta_{\mathcal{A}}(\omega_z)$ with $\eta_{\mathcal{A}}(\omega_z)\in H_{\mathcal{A}}(l+(m-1))$ and $b_z\in\mathbb{C}$. Write $\omega_{i_2}=y_{i_2}\omega'_{i_2}$ where $y_{i_2}\in\Sigma$ and $|\omega'_{i_2}|=l+(m-1)<l+m$. By the induction hypothesis, $$\eta_ {\mathcal{A}}(\omega'_{i_2})\in\mathcal{V}_{\mathcal{A}}(l)=
 \mathcal{V}_{\mathcal{A}}(l+(m-1)).$$ Thus,
 \begin{eqnarray}
 \label{eq21}
    \eta_{\mathcal{A}}(\omega'_{i_2})&=&\sum_kc_k\eta_{\mathcal{A}}(\omega'_{i_2,k})
    \qquad\mbox{($\omega'_{i_2,k}\in\Sigma^*$, $|\omega'_{i_2,k}|\leq l$ and $\,c_k\in\mathbb{C}$)}.
 \end{eqnarray}
 It follows that
 \begin{eqnarray*}
   \eta_{\mathcal{A}}(\omega_{i_2})&=&\eta_{\mathcal{A}}(y_{i_2}\omega'_{i_2})\\
                                   &=&A(y_{i_2})^{\dagger}\eta_{\mathcal{A}}
                                   (\omega'_{i_2})A(y_{i_2})
                                   \qquad\mbox{(by Eq.~(\ref{eq10}))}\\
                                   &=&A(y_{i_2})^{\dagger}\left(\sum_kc_k
                                   \eta_{\mathcal{A}}(\omega'_{i_2,k})\right)A(y_{i_2})
                                   \qquad\mbox{(by Eq.~(\ref{eq21}))}\\
                                   &=&\sum_kc_k(A(y_{i_2})^{\dagger}\eta_{\mathcal{A}}
                                   (\omega'_{i_2,k})A(y_{i_2}))\\
                                   &=&\sum_kc_k\eta_{\mathcal{A}}
                                   (y_{i_2}\omega'_{i_2,k})\qquad
                                   \mbox{(by Eq.~(\ref{eq10}))}.
 \end{eqnarray*}
Hence $\eta_{\mathcal{A}}(\omega_{i_2})\in\mathcal{V}_{\mathcal{A}}(l+1)$, as claimed. This completes the induction.\Q.E.D
\end{pf}

\begin{rmk}\label{rmk6}
  Further, if $\mathcal{A}_i=(Q_i,\{U_i(\sigma)\}_{\sigma\in\Sigma\cup\{\$\}},|\pi_i\rangle,
  \mathcal{O}_i)$, $i=1,2$, are two MM-1QFAs over $\Sigma$ and $\mathcal{A}=\mathcal{A}_1\oplus\mathcal{A}_2$ is their diagonal sum, then dim$\mathcal{V}_{\mathcal{A}}(i)\leq n_1^2+n_2^2$ for all $i\geq 0$, where $n_1=|Q_1|$ and $n_2=|Q_2|$. To see this, let
\begin{eqnarray*}
  \mathcal {B}&=&\{\,E_{ij}\,|\,1\leq i,j\leq n_1\,\}\cup\{\,E_{ij}\,|\,n_1+1\leq i,j\leq n_1+n_2\,\},
\end{eqnarray*}
where the elements in $\mathcal {B}$ are $(n_1+n_2)$-order matrices having only $1$ at the $(i,j)$ entry and $0$'s elsewhere.
Since, for all $\omega\in\Sigma^*$, 
\begin{eqnarray*}
   \eta_{\mathcal{A}}(\omega)&=&\left(
     \begin{array}{cc}
       \eta_{\mathcal{A}_1}(\omega) & 0 \\
       0 & \eta_{\mathcal{A}_2}(\omega) \\
     \end{array}
   \right),
\end{eqnarray*}
one can easy verify that
\begin{eqnarray*}
  \mathcal{V}_{\mathcal{A}}(i)&\subseteq& \mathrm{span}\{\mathcal {B}\}\qquad\mbox{( $\forall i\geq 0$ )}.
\end{eqnarray*}
This implies $\mathrm{dim}\mathcal{V}_{\mathcal{A}}(i)\leq n_1^2+n_2^2$ for all $i\geq 0$.  Hence, by replacing $\mathbb{M}_{|Q|}(\mathbb{C})$ with $\mathrm{span}\{\mathcal {B}\}$ in the proof of Lemma \ref{lem6}, we obtain $l<n_1^2+n_2^2$. 
\end{rmk}

The above remark shows the following

\begin{cor}\label{cor7}
Let $\mathcal{A}_i=(Q_i,\{U_i(\sigma)\}_{\sigma\in\Sigma\cup\{\$\}},|\pi_i\rangle,
\mathcal{O}_i)$, $i=1,2$, be two MM-1QFAs over $\Sigma$, and let $\mathcal{A}=\mathcal{A}_1\oplus\mathcal{A}_2$. Then there exists an integer $l<n_1^2+n_2^2$ (where $n_1=|Q_1|$ and $n_2=|Q_2|$) such that $\mathcal{V}_{\mathcal{A}}(l)=\mathcal{V}_{\mathcal{A}}(l+j)$ for all $j\geq 1$.\Q.E.D
\end{cor}

By virtue of Corollary \ref{cor7}, we prove the following theorem.
\begin{thm}
\label{thm8}
  Let $\mathcal{A}_i=(Q_i,\{U_i(\sigma)\}_{\sigma\in\Sigma\cup\{\$\}},|\pi_i\rangle,
  \mathcal{O}_i)$, $i=1,2$, be two MM-1QFAs over $\Sigma$, and let $\mathcal{A}=\mathcal{A}_1\oplus\mathcal{A}_2$. Then the unit vectors $|\varphi\rangle$ and $|\psi\rangle$, defined in Eqs.~(\ref{eq7}), are equivalent with respect to $\mathcal{A}$ if and only if they are $n_1^2+n_2^2-1$-equivalent with respect to $\mathcal{A}$, where $n_1$ and $n_2$ are the numbers of states in $\mathcal{A}_1$ and $\mathcal{A}_2$, respectively.
\end{thm}
\begin{pf}
 The ``only if" part is obvious. For the ``if" part, suppose $|\varphi\rangle$ and $|\psi\rangle$ are $n_1^2+n_2^2-1$-equivalent with respect to $\mathcal{A}$. Then for all $\omega=x_1x_2\cdots x_n\in\Sigma^{*}$ with $|\omega|<n_1^2+n_2^2-1$, Eq.~(\ref{eq8}) holds, i.e.,
\begin{eqnarray}
\label{eq22}
   \langle\varphi|\eta_{\mathcal{A}}(\omega)|\varphi\rangle&=&\langle\psi|
   \eta_{\mathcal{A}}(\omega)|\psi\rangle\qquad\mbox{($\forall\,\eta_{\mathcal{A}}
   (\omega)\in H_{\mathcal{A}}(n_1^2+n_2^2-1)$)}.
\end{eqnarray}
By Corollary \ref{cor7}, for all $\omega\in\Sigma^{*}$, we have $$\eta_{\mathcal{A}}(\omega)\in \mathcal{V}_{\mathcal{A}}(n_1^2+n_2^2-1)=\mathrm{span}\big\{H_{\mathcal{A}}
(n_1^2+n_2^2-1)\big\}.$$ Hence,
\begin{eqnarray}  \label{eq23}
  \eta_{\mathcal{A}}(\omega)&=&\sum_ia_i\,\eta_{\mathcal{A}}(\omega_i)\qquad
  \mbox{($\eta_{\mathcal{A}}(\omega_i)\in H_{\mathcal{A}}\big(n_1^2+n_2^2-1\big)$)}
\end{eqnarray}
for some $a_i\in\mathbb{C}$. It follows that
\begin{eqnarray*}
  \langle\varphi|\,\eta_{\mathcal{A}}(\omega)\,|\varphi\rangle&=&\langle\varphi|\,
\left(\sum_ia_i\,\eta_{\mathcal{A}}(\omega_i)\right)\,|\varphi\rangle
  \qquad\mbox{(by Eq.~(\ref{eq23}))}\\
  &=&\sum_ia_i\,\Big(\langle\varphi|\,\eta_{\mathcal{A}}(\omega_i)\,
|\varphi\rangle\Big)\qquad\mbox{($\eta_{\mathcal{A}}(\omega_i)\in H_{\mathcal{A}}\big(n_1^2+n_2^2-1\big)$)}\\
  &=&\sum_ia_i\,\Big(\langle\psi|\,\eta_{\mathcal{A}}(\omega_i)\,|\psi\rangle\Big)
  \qquad\mbox{(by Eq.~(\ref{eq22}))}\\
  &=&\langle\psi|\,\eta_{\mathcal{A}}(\omega)\,|\psi\rangle.
\end{eqnarray*}
Thus Eq.~(\ref{eq8}) holds for all $\omega\in\Sigma^{*}$, so $|\varphi\rangle$ and $|\psi\rangle$ are equivalent with respect to $\mathcal{A}$.\Q.E.D
\end{pf}

\par
Now we can present the proof of Theorem \ref{thm1}.\\

\par
\noindent{\bf Proof of Theorem \ref{thm1}.} By Theorem \ref{thm4}, it suffices to show that $\mathcal{A}_1$ and $\mathcal{A}_2$ are $\beta$-equivalent if and only if they are $(n_1^2+n_2^2-1)$-$\beta$-equivalent.

\par
  Since the ``only if" direction is obvious, we only need to prove the ``if" direction. Let $\mathcal{A}=\mathcal{A}_1\oplus\mathcal{A}_2$. By Remark \ref{rmk3},
\begin{eqnarray}\label{eq24}
  \mathcal{F}_{\mathcal{A}_1}(\omega)&=&\langle\varphi|\,\eta_{\mathcal{A}}(\omega)\,
  |\varphi\rangle
\end{eqnarray}
and
\begin{eqnarray}\label{eq25}
  \mathcal{F}_{\mathcal{A}_2}(\omega)&=&\langle\psi|\,\eta_{\mathcal{A}}(\omega)\,
  |\psi\rangle
\end{eqnarray}
for all $\omega\in\Sigma^{*}$ (where $|\varphi\rangle$ and $|\psi\rangle$ are defined in Eqs.~(\ref{eq7})).

Suppose  $\mathcal{A}_1$ and $\mathcal{A}_2$ are $(n_1^2+n_2^2-1)$-$\beta$-equivalent. Then
\begin{eqnarray}\label{eq26}
  \mathcal{F}_{\mathcal{A}_1}(\omega)&=&\mathcal{F}_{\mathcal{A}_2}(\omega)
\end{eqnarray}
for all $\omega\in\Sigma^{*}$ with $|\omega|<n_2^2+n_2^2-1$.
Combining  Eq.~(\ref{eq24}), Eq.~(\ref{eq25}), and Eq.~(\ref{eq26}) yields
\begin{eqnarray*}
  \langle\varphi|\,\eta_{\mathcal{A}}(\omega)\,|\varphi\rangle&=&\langle\psi|\,
  \eta_{\mathcal{A}}(\omega)\,|\psi\rangle
  \qquad\mbox{($|\omega|<n_1^2+n_2^2-1$)}.
\end{eqnarray*}
This means that $|\varphi\rangle$ and $|\psi\rangle$ are $n_1^2+n_2^2-1$-equivalent with respect to $\mathcal{A}$. By Theorem \ref{thm8}, they are equivalent with respect to $\mathcal{A}$, so $$\langle\varphi|\,\eta_{\mathcal{A}}(\omega)\,|\varphi\rangle=\langle\psi|\,
\eta_{\mathcal{A}}(\omega)\,|\psi\rangle$$ for all $\omega\in\Sigma^{*}$, i.e., $\mathcal{F}_{\mathcal{A}_1}(\omega)=\mathcal{F}_{\mathcal{A}_2}(\omega)$ for all $\omega\in\Sigma^{*}$. Hence $\mathcal{A}_1$ and $\mathcal{A}_2$ are $\beta$-equivalent.\Q.E.D

One may argue that the improvement from $3n_1^2+3n_2^2-1$ to $n_1^2+n_2^2-1$ is not essential, since  both bounds are quadratic. We conjecture that the upper-bound $n_1^2+n_2^2-1$ cannot be further improved to a linear bound. However, we have no ability to prove this.

\section{Proof of Theorem \ref{thm2}}
\label{sec:proof_of_theorem2}

In this section, we investigate the equivalence problem for E-1QFAs. For convenience, we adopt the following notations.

\par
For any $i\geq 0$, let  $H_{\mathcal{A}}(i)$ denote the set
$$\{\theta_{\mathcal{A}}(\omega)|\omega\in\Sigma^*,|\omega|\leq i\},$$ let $V_{\mathcal{A}}(i)$ be the vector space spanned by $H_{\mathcal{A}}(i)$, let $K_{\mathcal{A}}(i)$ denote the set
$\{\vartheta_{\mathcal{A}}(\omega)|\omega\in\Sigma^*,|\omega|\leq i\}$, and let $\mathcal{S}_{\mathcal{A}}(i)$ be the vector space spanned by $K_{\mathcal{A}}(i)$. The following inclusions are immediate:
\begin{eqnarray*}
H_{\mathcal{A}}(i)\subseteq H_{\mathcal{A}}(i+1),\qquad
V_{\mathcal{A}}(i)\subseteq V_{\mathcal{A}}(i+1)\\
 K_{\mathcal{A}}(i)\subseteq K_{\mathcal{A}}(i+1),\qquad
 \mathcal{S}_{\mathcal{A}}(i)\subseteq\mathcal{S}_{\mathcal{A}}(i+1).
\end{eqnarray*}

\begin{lem}\label{lem9}
Let $\mathcal{A}_i=(Q_i,Q_{acc,i},Q_{rej,i},\{\mathcal{U}_{\sigma}^{(i)}\}_{\sigma\in
\Sigma\cup\{\#,\$\}},\rho_i,\mathcal{O}_i)$, $i=1,2$, be two E-1QFAs over $\Sigma$, and let $\mathcal{A}=\mathcal{A}_1\oplus\mathcal{A}_2$. Then there exists an integer $l<n_1^2+n_2^2$, where $n_1=|Q_1|$ and $n_2=|Q_2|$, such that $\mathcal{S}_{\mathcal{A}}(l)=\mathcal{S}_{\mathcal{A}}(l+j)$ for all $j\geq 1$.
\end{lem}
\begin{pf}
The proof is similar to that of Lemma \ref{lem6}. First, note that if $\mathcal{A}_i$ ($i=1,2$) are two E-1QFAs over $\Sigma$ and $\mathcal{A}=\mathcal{A}_1\oplus\mathcal{A}_2$, then
 \begin{eqnarray*}
   \vartheta_{\mathcal{A}}(\omega)&=&\left(
                                       \begin{array}{cc}
                                         \vartheta_{\mathcal{A}_1}(\omega) & 0 \\
                                         0 & \vartheta_{\mathcal{A}_2}(\omega) \\
                                       \end{array}
                                     \right)
 \end{eqnarray*}
 for all $\omega\in\Sigma^*$.
 Hence, by an argument similar to that in Remark \ref{rmk6}, we have dim$\mathcal{S}_{\mathcal{A}}(i)\leq n_1^2+n_2^2$ for all $i\geq 0$.

Using the same reasoning as in the proof of Lemma \ref{lem6}, there exists an integer $l<n_1^2+n_2^2$ such that $\mathcal{S}_{\mathcal{A}}(l)=\mathcal{S}_{\mathcal{A}}(l+1)$.

  \par
  Next, we show by induction on $j$ that $\mathcal{S}_{\mathcal{A}}(l)=\mathcal{S}_{\mathcal{A}}(l+j)$ for all $j\geq 1$. The case $j=1$ has already been established. Assume the statement holds for all $j<m$ ($m>1$) and consider $j=m$. Since $\mathcal{S}_{\mathcal{A}}(l+m)=\mathrm{span}\{K_{\mathcal{A}}(l+m)\}$ and $$K_{\mathcal{A}}(l+m)=K_{\mathcal{A}}(l+(m-1))\cup
  \{\vartheta_{\mathcal{A}}(\omega)|\omega\in\Sigma^*,|\omega|=l+m\},$$ any $\vartheta\in\mathcal{S}_{\mathcal{A}}(l+m)$ can be written as
  \begin{eqnarray*}
    \vartheta&=&\sum_{i_1}a_{i_1}\,\vartheta_{\mathcal{A}}(\omega_{i_1})+
    \sum_{i_2}a_{i_2}\,\vartheta_{\mathcal{A}}(\omega_{i_2})
  \end{eqnarray*}
  where $\vartheta_{\mathcal{A}}(\omega_{i_1})\in K_{\mathcal{A}}(l+(m-1))$ and $\vartheta_{\mathcal{A}}(\omega_{i_2})\in\{\vartheta_{\mathcal{A}}(\omega)|
  \omega\in\Sigma^*,|\omega|=l+m\}$. It suffices to show that
  \begin{eqnarray}\label{eq28}
  \sum_{i_2}a_{i_2}\,\vartheta_{\mathcal{A}}(\omega_{i_2})&\in&
  \mathcal{S}_{\mathcal{A}}(l+(m-1)).
  \end{eqnarray}

  \par
  Note that $|\omega_{i_2}|=l+m$. Write $\omega_{i_2}=yx_1x_2\cdots x_{l+(m-1)}$ with $y\in\Sigma$. Then
  \begin{eqnarray*}
   \vartheta_{\mathcal{A}}(\omega_{i_2})&=&\sum_{i_y}B_{i_y}^{\dagger}
   \vartheta_{\mathcal{A}}(x_1x_2\cdots x_{l+(m-1)})B_{i_y}\qquad\mbox{(by Eq.~(\ref{eq19}))}\\
   &&\mbox{(by induction hypothesis, we have)}\\
   &=&\sum_{i_y}B_{i_y}^{\dagger}\left(\sum_za_z\vartheta_{\mathcal{A}}
   (\omega_z)\right)B_{i_y}\qquad\mbox{($\vartheta_{\mathcal{A}}(\omega_z)\in K_{\mathcal{A}}(l)$)}\\
   &=&\sum_za_z\left(\sum_{i_y}B_{i_y}^{\dagger}\vartheta_{\mathcal{A}}
   (\omega_z)B_{i_y}\right)\\
   &=&\sum_za_z\,\vartheta_{\mathcal{A}}(y\omega_z)\qquad\mbox{(by Eq.~(\ref{eq19}))}
  \end{eqnarray*}
  with $|y\omega_z|\leq l+1$ and $a_z\in\mathbb{C}$. This shows that $\vartheta\in\mathcal{S}_{\mathcal{A}}(l+m)$, completing the induction. \Q.E.D
\end{pf}

\par
Now we prove the following lemma.

\begin{lem}\label{lem10}
Let $\mathcal{A}_i=(Q_i,Q_{acc,i},Q_{rej,i},
\{\mathcal{U}_{\sigma}^{(i)}\}_{\sigma\in\Sigma\cup\{\#,\$\}},\rho_i,\mathcal{O}_i)$, $i=1,2$,
be two E-1QFAs over $\Sigma$, and let $\mathcal{A}=\mathcal{A}_1\oplus\mathcal{A}_2$. Then $$V_{\mathcal{A}}(n_1^2+n_2^2-1)=V_{\mathcal{A}}((n_1^2+n_2^2-1)+j)$$ for all $j\geq 1$.
\end{lem}
\begin{pf}
 For any $\omega\in\Sigma^*$ with $|\omega|=(n_1^2+n_2^2-1)+j$, we have
 \begin{eqnarray*}
   \theta_{\mathcal{A}}(\omega)&=&\sum_{i_{x_0}}B_{i_{x_0}}^{\dagger}
   \vartheta_{\mathcal{A}}(\omega)B_{i_{x_0}}\qquad\mbox{(by Eq.~(\ref{eq15}))}\\
   &&\mbox{(by Lemma \ref{lem9}, we have)}\\
   &=&\sum_{i_{x_0}}B_{i_{x_0}}^{\dagger}\left(\sum_za_z\,\vartheta_{\mathcal{A}}
   (\omega_z)\right)B_{i_{x_0}}\qquad\mbox{( $\vartheta_{\mathcal{A}}(\omega_z)\in K_{\mathcal{A}}(n_1^2+n_2^2-1)$ )}\\
   &=&\sum_za_z\left(\sum_{i_{x_0}}B_{i_{x_0}}^{\dagger}\vartheta_{\mathcal{A}}
   (\omega_z)B_{i_{x_0}}\right)\\
   &=&\sum_za_z\,\theta_{\mathcal{A}}(\omega_z)\qquad\mbox{(by Eq.~(\ref{eq15}))}
 \end{eqnarray*}
 where $|\omega_z|\leq n_1^2+n_2^2-1$ and $a_z\in\mathbb{C}$. Hence $$V_{\mathcal{A}}((n_1^2+n_2^2-1)+j)=V_{\mathcal{A}}(n_1^2+n_2^2-1).$$ Since the argument holds for all $j\geq 1$, the lemma follows. \Q.E.D
\end{pf}

\begin{rmk}
 Note that the proof of Lemma \ref{lem10} relies on Lemma \ref{lem9} because an E-1QFA has the left end-mark `\#', which prevents a direct proof. This is also why $\theta_{\mathcal{A}}(\omega)$ is expressed in the form given by  Eq.~(\ref{eq15}).
\end{rmk}

\par
The proofs of the following theorem and of Theorem \ref{thm2} are similar to those of Theorem \ref{thm8} and Theorem \ref{thm1}, respectively. For self-containedness, we present them in detail.

\begin{thm}\label{thm11}
 Let $\mathcal{A}_i=(Q_i,Q_{acc,i},Q_{rej,i},\{\mathcal{U}_{\sigma}^{(i)}\}_{\sigma
 \in\Sigma\cup\{\#,\$\}},\rho_i,\mathcal{O}_i)$, $i=1,2$, be two E-1QFAs over $\Sigma$, and let $\mathcal{A}=\mathcal{A}_1\oplus\mathcal{A}_2$. Then the density matrices $\varphi$ and $\psi$, defined in Eqs.~(\ref{eq16}), are equivalent with respect to $\mathcal{A}$ if and only if they are $n_1^2+n_2^2-1$-equivalent with respect to $\mathcal{A}$, where $n_1$ and $n_2$ are the numbers of states in $\mathcal{A}_1$ and $\mathcal{A}_2$, respectively.
\end{thm}

\begin{pf}
The ``only if" part is trivial. For the ``if" part, assume that $\varphi$ and $\psi$ are $n_1^2+n_2^2-1$-equivalent. Then for all $\omega\in\Sigma^*$ with $|\omega|\leq n_1^2+n_2^2-1$, Eq.~(\ref{eq17}) holds:
\begin{eqnarray}\label{eq29}
 (\langle q_0^{(1)}|,{\bf 0})\theta_{\mathcal{A}}(\omega)
 \left(
    \begin{array}{c}
    |q_0^{(1)}\rangle \\
    {\bf 0} \\
   \end{array}
 \right)&=&({\bf 0},\langle q_0^{(2)}|)\theta_{\mathcal{A}}(\omega)
 \left(
    \begin{array}{c}
      {\bf 0} \\
      |q_0^{(2)}\rangle \\
      \end{array}
      \right)
   \end{eqnarray}
 for all $\theta_{\mathcal{A}}(\omega)\in V_{\mathcal{A}}(n_1^2+n_2^2-1)$.

 \par
 By Lemma \ref{lem10}, for all $\omega\in\Sigma^*$ we have
 \begin{eqnarray}\label{eq30}
 \theta_{\mathcal{A}}(\omega)&=&\sum_ia_i\theta_{\mathcal{A}}(\omega_i)
 \qquad\mbox{($\theta_{\mathcal{A}}(\omega_i)\in H_{\mathcal{A}}(n_1^2+n_2^2-1), a_i\in\mathbb{C}$)}.
 \end{eqnarray}
Thus,
\begin{eqnarray*}
(\langle q_0^{(1)}|,{\bf 0})\theta_{\mathcal{A}}(\omega)
\left(
     \begin{array}{c}
     |q_0^{(1)}\rangle \\
     {\bf 0} \\
     \end{array}
     \right)
   &=&(\langle q_0^{(1)}|,{\bf 0})\left(\sum_ia_i\theta_{\mathcal{A}}(\omega_i)\right)
   \left(
     \begin{array}{c}
     |q_0^{(1)}\rangle \\
     {\bf 0} \\
     \end{array}
     \right)\qquad\mbox{(by Eq.~(\ref{eq30}))}\\
   &=&\sum_ia_i\left((\langle q_0^{(1)}|,{\bf 0})\theta_{\mathcal{A}}(\omega_i)
   \left(
   \begin{array}{c}
      |q_0^{(1)}\rangle \\
      {\bf 0} \\
      \end{array}
      \right)
   \right)\\
   &=&
   \sum_ia_i\left(({\bf 0},\langle q_0^{(2)}|)\theta_{\mathcal{A}}(\omega_i)
   \left(
     \begin{array}{c}
     {\bf 0} \\
     |q_0^{(2)}\rangle \\
     \end{array}
     \right)\right)\qquad\mbox{(by Eq.~(\ref{eq29}))}\\
     &=&({\bf 0},\langle q_0^{(2)}|)\left(\sum_ia_i\theta_{\mathcal{A}}(\omega_i)\right)
     \left(
     \begin{array}{c}
     {\bf 0} \\
       |q_0^{(2)}\rangle \\
       \end{array}
     \right)\\
     &=&({\bf 0},\langle q_0^{(2)}|)\theta_{\mathcal{A}}(\omega)
     \left(
        \begin{array}{c}
        {\bf 0} \\
        |q_0^{(2)}\rangle \\
        \end{array}
        \right)\qquad\mbox{(by Eq.~(\ref{eq30}))}.
   \end{eqnarray*}
Hence Eq.~(\ref{eq17}) holds for all $\omega\in\Sigma^*$. By Definition \ref{df10}, $\varphi$ and $\psi$ are equivalent with respect to $\mathcal{A}$.\Q.E.D
\end{pf}

\par
Finally, we present the proof of Theorem \ref{thm2}.\\

\par
\noindent{\bf Proof of Theorem \ref{thm2}.} By Theorem \ref{thm5}, it suffices to show that $\mathcal{A}_1$ and $\mathcal{A}_2$ are $\beta$-equivalent if and only if they are $(n_1^2+n_2^2-1)$-$\beta$-equivalent.

\par
The ``only if" direction is clear. Assume $\mathcal{A}_1$ and $\mathcal{A}_2$ are $(n_1^2+n_2^2-1)-\beta$-equivalent. Let $\mathcal{A}=\mathcal{A}_1\oplus\mathcal{A}_2$. Then for all $\omega\in\Sigma^*$ with $|\omega|\leq n_1^2+n_2^2-1$, 
 \begin{eqnarray}\label{eq31}
   \mathcal{F}_{\mathcal{A}_1}(\omega)&=&\mathcal{F}_{\mathcal{A}_2}(\omega)
   \qquad\mbox{( $|\omega|\leq n_1^2+n_2^2-1$ )}
 \end{eqnarray}
 By Remark \ref{rmk5},
 \begin{eqnarray}
   \mathcal{F}_{\mathcal{A}_1}(\omega)&=&(\langle q_0^{(1)}|,{\bf 0})\,\,\theta_{\mathcal{A}}(\omega)\,\,\left(
                                                                \begin{array}{c}
                                                                  |q_0^{(1)}\rangle \\
                                                                  {\bf 0} \\
                                                                \end{array}
                                                              \right)\label{eq32}\\
   \mathcal{F}_{\mathcal{A}_2}(\omega)&=&({\bf 0},\langle q_0^{(2)}|)\,\,\theta_{\mathcal{A}}(\omega)\,\,\left(
                                                               \begin{array}{c}
                                                                 {\bf 0} \\
                                                                 |q_0^{(2)}\rangle \\
                                                               \end{array}
                                                             \right)\label{eq33}
 \end{eqnarray}
 Eqs.~(\ref{eq31}), (\ref{eq32}), and (\ref{eq33}) imply that $\varphi$ and $\psi$ are $n_1^2+n_2^2-1$-equivalent with respect to $\mathcal{A}$. By Theorem \ref{thm11}, $\varphi$ and $\psi$ are equivalent with respect to $\mathcal{A}$, so $$\mathcal{F}_{\mathcal{A}_1}(\omega)=\mathcal{F}_{\mathcal{A}_2}(\omega)$$ for all $\omega\in\Sigma^*$. Therefore, $\mathcal{A}_1$ and $\mathcal{A}_2$ are $\beta$-equivalent. Theorem \ref{thm2} follows.\Q.E.D

\section{Conclusions}
\label{sec:conclusions}

In this paper, we have shown that two MM-1QFAs $\mathcal{A}_1$ and $\mathcal{A}_2$ over the same alphabet $\Sigma$ are equivalent if and only if they are $(n_1^2+n_2^2-1)$-equivalent. Our result indicates that the upper-bound for the equivalence problem of MM-1QFAs is independent of the numbers of states in the minimal DFA that recognizes the regular language $g^*a\{a,r,g\}^*$. The approach used in this paper is similar to the work of Carlyle \cite{4}. Moreover, compared with \cite{26}, the reader may find that the approach used in this paper is much simpler, more direct, and more elegant.

As an application of this approach, we use it to solve the equivalence problem for E-1QFAs, which has not been answered previously, by proving Theorem \ref{thm2}.

As mentioned earlier, from the algebraic point of view, the concept of ``equivalence" provides a classification of the elements in the set of MM-1QFAs over the same alphabet. Let $\mathcal{A}$ be an MM-1QFA over $\Sigma$, and let $\widetilde{\mathcal{A}}$ denote the set of all MM-1QFAs over $\Sigma$ that are equivalent to $\mathcal{A}$. Then a natural question arises: Does there exists an MM-1QFA $\mathcal{A}'\in\widetilde{\mathcal{A}}$ with the smallest (minimal) numbers of basic states? If such an element exists, how can it be construct? It is our future work to consider these interesting and more challenging problems.

\section*{Acknowledgements}
The author is indebted to the anonymous referees and the editors, especially one of the referees, for their invaluable comments and suggestions, which helped to considerably improve the quality and presentation of this paper.


%%\bibliographystyle{elsarticle-num}
%%\bibliography{<your-bib-database>}

\end{document}